\documentclass[twocolumn, prl, longbibliography]{revtex4-1}
\usepackage{amsmath, amssymb, color, graphicx, tikz}
\definecolor{linkcolor}{rgb}{0,0,0.6} 
\usepackage[pdftex,colorlinks=true,
	pdfstartview = FitV,
	linkcolor    = linkcolor,
	citecolor    = linkcolor,
	urlcolor     = linkcolor,	
	hyperindex   = true,
	hyperfigures = false]{hyperref}

\definecolor{lgreen} {RGB}{180,210,100}
\definecolor{dblue}  {RGB}{20,66,129}
\definecolor{jblue}  {RGB}{20,50,100}
\definecolor{nblue}  {RGB}{0,120,200}
\definecolor{dgreen} {RGB}{78,138,21}
\definecolor{ngreen} {RGB}{98,158,31}
\definecolor{lred}   {RGB}{220,0,0}
\definecolor{nred}   {RGB}{224,0,0}
\definecolor{norange}{RGB}{230,120,20}
\definecolor{nyellow}{RGB}{255,221,0}

\begin{document}

\title{Thermodynamic cycles with active matter}

\author{Timothy Ekeh}
\author{Michael E. Cates}
\author{\'Etienne Fodor}
\affiliation{DAMTP, Centre for Mathematical Sciences, University of Cambridge, Wilberforce Road, Cambridge CB3 0WA, UK}

\begin{abstract}
Active matter constantly dissipates energy to power the self-propulsion of its microscopic constituents. This opens the door to designing innovative cyclic engines without any equilibrium equivalent. We offer a consistent thermodynamic framework to characterize and optimize the performances of such cycles. Based on a minimal model, we put forward a protocol which extracts work by controlling only the properties of the confining walls at boundaries, and we rationalize the transitions between optimal cycles. We show that the corresponding power and efficiency are generally proportional, so that they reach their maximum values at the same cycle time in contrast with thermal cycles, and we provide a generic relation constraining the fluctuations of the power.
\end{abstract}

\maketitle


The properties of thermal engines, which operate typically with cycles of temperature and volume, are well described within the framework of standard thermodynamics. Simple protocols, such as the Carnot and the Stirling cycles, provide an intuitive understanding of the minimal rules required to extract maximal work and dissipate minimal heat out of ideal fluids~\cite{Carnot}. As such, they still serve today as insightful references to develop optimal cycles in more realistic settings. More recently, they have also been used to test the concepts of stochastic thermodynamics in experiments where fluctuations cannot be neglected~\cite{Blickle2011, Sood2016, Martinez2016}.

During the last decades, active matter has emerged as an important class of nonequilibrium systems where particles extract energy from their environment to power a directed motion~\cite{Marchetti2013, Bechinger2016, Marchetti2018}. Swarms of bacteria~\cite{Kessler2004, Goldstein2007, Aranson2012} and assemblies of Janus colloids in a fuel bath~\cite{Bechinger2013, Palacci2013} are typical examples where the microscopic dissipation controls the macroscopic fluid properties. A number of theoretical works have strived to build a thermodynamic approach to rationalize these properties by analogy with equilibrium~\cite{Tailleur2008, Fily2012, Speck2014, Maggi2015, Nardini2016, Nardini2017, Cugliandolo2018, Nemoto2019}. In minimal models where the solvent only provides passive friction and momentum is not conserved, the pressure is not an equation of state, at variance with equilibrium, since it generally depends on the properties of the wall used to measure it~\cite{Brady2014, Solon2015a, Solon2015b}. In these models, a definition of chemical potential has also been proposed which highlights again the limitations of equilibrium analogies~\cite{Paliwal2018, Guioth2019}.

In thermal systems, work can be extracted from cyclic protocols only by establishing a heat flow in the system, for instance with a periodic change of temperature. In active matter, heat flows are already present at fixed temperature due to individual self-propulsion. Autonomous engines can then be designed by promoting the current of asymmetric obstacles~\cite{Sokolov2010, Leonardo2010, Leonardo2017} and extracting work with an external load~\cite{Pietzonka2019}. In principle, monothermal cycles can also extract work out of active matter in the case where macroscopic currents are absent. It remains to determine how to exploit properly nonequilibrium properties in active matter to design such cycles, and how to build a generic approach to quantify, compare and optimize systematically their performances.


In this paper, we provide a thermodynamic framework to investigate systematically the performances of monothermal cyclic engines operating with active matter. As a popular model of active fluids, we consider a set of $N$ independent Active Brownian Particles (ABPs) in two dimensions~\cite{Fily2012}. They are subject to external confining and aligning potentials, respectively denoted by $u_{\rm t}$ and $u_{\rm r}$. Neglecting particle interactions, the dynamics of position ${\bf r}_i$ and orientation $\theta_i$ reads
\begin{equation}\label{eq:dyn}
	\begin{aligned}
		\dot{\bf r}_i &= v {\bf e}_i  - \mu_{\rm t} \nabla_i u_{\rm t} + \sqrt{2D_{\rm t}} {\boldsymbol\xi}_i ,
		\\
		\dot\theta_i &= - \mu_{\rm r} \partial_{\theta_i} u_{\rm r} + \sqrt{2D_{\rm r}}\eta_i ,
	\end{aligned}
\end{equation}
where $v$ is the self-propulsion speed, ${\bf e}_i = (\cos\theta_i, \sin\theta_i)$ the orientation vector, and $\{{\boldsymbol\xi}_i, \eta_i\}$ a set of uncorrelated Gaussian white noises with zero mean and unit variance. The translational and rotational mobilities $\{\mu_{\rm t}, \mu_{\rm r}\}$ are independent in general, and so are the translational and rotational diffusion constants $\{D_{\rm t}, D_{\rm r}\}$.

\begin{figure}[b]
	\centering
	\includegraphics[width=\linewidth, trim=1.8cm 22.4cm 11cm 1.9cm, clip=true]{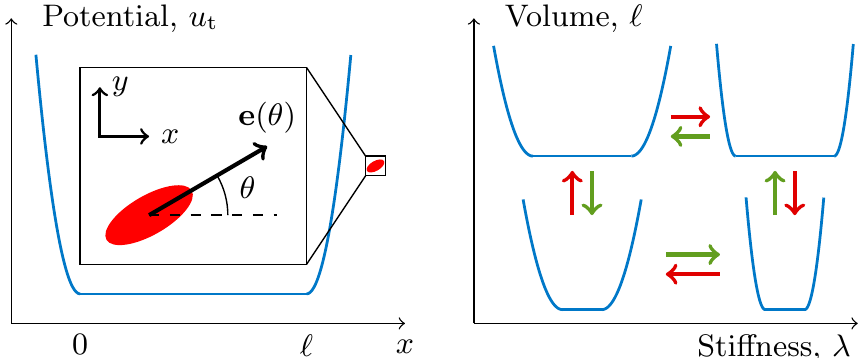}
	\caption{\label{fig1}
		Schematic illustration of the active engine.
		(Left)~Elliptical active particles are confined between two parallel walls separated by a distance $\ell$ with stiffness $\lambda$.
		(Right)~The cycle of volume and stiffness, operating either clockwise or counter-clockwise, extracts work by controlling only confining walls.
	}
\end{figure}

To extract work, we suppose that the operator can modify externally a series of parameters $\{\alpha_1, \dots, \alpha_n\}$ which control the shape of the potentials $u_{\rm t}$ and $u_{\rm r}$, see Fig.~\ref{fig1}. The tools of stochastic thermodynamics, introduced originally for thermal systems~\cite{Sekimoto1998, Seifert2012} and later extended to active ones~\cite{Speck2016, Ahmed2016, Nardini2016, Cagnetta2017, Nardini2017, Seifert2018, Bo2019, Suri2019}, allow us to identify the average incremental work associated with an infinitesimal variation of an arbitrary number of parameters as $\delta{\cal W} = N\sum_n \langle\partial_{\alpha_n} u_{\rm tot}\rangle {\rm d}\alpha_n $, where $u_{\rm tot}=u_{\rm t}+u_{\rm r}$ and $\langle\cdot\rangle$ is the average with respect to noise realisations. For quasistatic protocols, it is sufficient to evaluate averages in steady state at fixed $\alpha_n$ denoted by $\langle\cdot\rangle_{\rm s}$. Considering a cyclic protocol $\partial\Sigma$ which encloses the surface $\Sigma$ in the space of two independent parameters, the average quasistatic work ${\cal W}_{\rm qs}$ then reduces to
\begin{equation}\label{eq:work_def}
	{\cal W}_{\rm qs} = N\oint_{\partial\Sigma} \bigg[ \bigg\langle\frac{\partial u_{\rm tot}}{\partial\alpha_1}\bigg\rangle_{\rm s}{\rm d}\alpha_1 + \bigg\langle\frac{\partial u_{\rm tot}}{\partial\alpha_2}\bigg\rangle_{\rm s}{\rm d}\alpha_2 \bigg] ,
\end{equation}
which can also be written using Green's theorem as
\begin{equation}\label{eq:work_green}
	\begin{aligned}
		{\cal W}_{\rm qs} &= \pm \,N\iint_\Sigma w(\alpha_1, \alpha_2) \,{\rm d}\alpha_1{\rm d}\alpha_2 ,
		\\
		w(\alpha_1, \alpha_2) &= \frac{\partial}{\partial\alpha_2}\bigg\langle\frac{\partial u_{\rm tot}}{\partial\alpha_1}\bigg\rangle_{\rm s} - \frac{\partial}{\partial\alpha_1}\bigg\langle\frac{\partial u_{\rm tot}}{\partial\alpha_2}\bigg\rangle_{\rm s} ,
	\end{aligned}
\end{equation}
where $+$ and $-$ signs respectively refer to clockwise and counter-clockwise protocols in the $\{\alpha_1,\alpha_2\}$ plane. With our convention, the cycle extracts work from the system whenever ${\cal W}_{\rm qs}<0$.

At equilibrium ($v=0$ and $u_{\rm t}=u_{\rm r}=u$), the weight of configurations follows the Boltzmann factor ${\rm e}^{-u/T}$. The temperature $T=D_{\rm t}/\mu_{\rm t}=D_{\rm r}/\mu_{\rm r}$ enforces a constraint between mobilities and diffusion constants, and the averages in~\eqref{eq:work_green} are written in terms of the free energy ${\cal F} = -N T\ln\big[ \int {\rm e}^{-u/T} {\rm d}{\bf r}{\rm d}\theta\big] $ as $\langle\partial_{\alpha_n} u\rangle_{\rm s} = \partial_{\alpha_n}\cal F$. Then, the quasistatic work always vanishes independently of the cycle details, as expected from standard thermodynamics. For generic active fluids, the steady state is no longer given by the Boltzmann distribution, as a consequence of the breakdown of detailed balance~\cite{Solon2015, Maggi2015, Nardini2016}. Hence, work can now potentially be extracted by tuning only the external parameters $\{\alpha_1,\alpha_2\}$ without varying any internal parameter of the dynamics.

In what follows, we consider that the volume of the system and the stiffness of confining walls change periodically, as shown in Fig.~\ref{fig1}. This cycle illustrates how controlling active systems only at boundaries is actually sufficient to extract work, without changing any property of the microscopic constituents, at variance with thermal cycles. We first compute the average work for quasistatic protocols. This sheds light on a transition of the appropriate cycle direction to extract work, either clockwise and counter-clockwise, recapitulated in a phase diagram in terms of microscopic parameters. For finite cycle time, we then provide a generic relation between the average and the variance of extracted power, and we show that the cycle efficiency, defined in terms of work and heat, is proportional to the average power.


The active particles are confined along $\hat x$ by two parallel walls with translational invariance along $\hat y$. Inspired by a recent work~\cite{Solon2015a}, we take the confining and aligning potentials as $u_{\rm t} = (\lambda/2) \big[(x-\ell)^2 H(x-\ell) + x^2H(-x)\big]$ and $u_{\rm r} = (\lambda\kappa/2)\cos(2\theta) \big[H(x-\ell) + H(-x)\big]$, where $H$ is the Heaviside step function. The control parameters are the distance between the walls $\ell$, which sets the volume of the system, and the stiffness of the walls $\lambda$. The parameter $\kappa$, kept constant throughout the protocol, determines the tendency of particles to align parallel to the wall. For elliptical particles of axial dimensions $\{a,b\}$, as shown in Fig.~\ref{fig1}, $\kappa$ is proportional to the anisotropy $a^2-b^2$, and it vanishes for isotropic particles ($a=b$)~\cite{Solon2015a}. Note that the stiffness sets the amplitudes of both confining and aligning potentials.

With these settings, the average quasistatic work~\eqref{eq:work_def} extracted from the cycle of volume and stiffness $\partial\Sigma$ reads
\begin{equation}\label{eq:work_cycle}
	{\cal W}_{\rm qs} = \oint_{\partial\Sigma} \bigg[ - P {\rm d}\ell + N \langle u_{\rm tot}\rangle_{\rm s} \frac{{\rm d}\lambda}{\lambda}\bigg] ,
\end{equation}
where we have introduced the pressure exerted on the right wall $P = - N\langle \partial_{\ell} u_{\rm t} \rangle_{\rm s} = N\lambda\langle (x-\ell) H(x-\ell) \rangle_{\rm s}$~\cite{Solon2015a, Solon2015b}. Though $P$ is defined independently of the torque exerted by the wall, its explicit expression depends on $\kappa$ in general. The first term in~\eqref{eq:work_cycle}, which embodies the work extracted by compressing and expanding the system, has a similar form as in equilibrium except that the pressure now potentially differs for active fluids. The second one quantifies the work required to stiffen and soften the wall.

It is well documented that active particles accumulate at the walls for small angular diffusion $D_{\rm r}\ll\lambda\mu_{\rm t}$~\cite{Maggi2015, Solon2015, Bechinger2016}, thus affecting the density profile beyond the wall regions. To evaluate explicitly $P$ and $\langle u_{\rm tot} \rangle_{\rm s}$, we focus on the opposite regime where the distribution of position and orientation is flat between the walls. Since the confining potential $u_{\rm t}$ is soft, particles can penetrate the wall and thereby deplete the bulk: The bulk density $\rho$ varies when changing either volume or stiffness. To account for this effect, we approximate the distribution in the wall regions by a Boltzmann factor with effective temperature $D_{\rm t}(1+{\rm Pe})/\mu_{\rm t}$, where ${\rm Pe} = v^2/(2D_{\rm t}D_{\rm r})$ is the P\'eclet number, leading to
\begin{equation}\label{eq:rho}
	\rho(\ell,\lambda) = \frac{N}{\ell + \sqrt{2\pi D_{\rm t}(1 +{\rm Pe}) / (\lambda\mu_{\rm t})}} .
\end{equation}
In practice, the regime where the wall penetration provides a significant contribution to the bulk density $\rho$ is consistent with the effective temperature approximation~\cite{Supplemental}. Importantly, we only use this approximation when renormalizing the bulk density as in~\eqref{eq:rho}.

\begin{figure*}
	\centering
	\includegraphics[width=\linewidth]{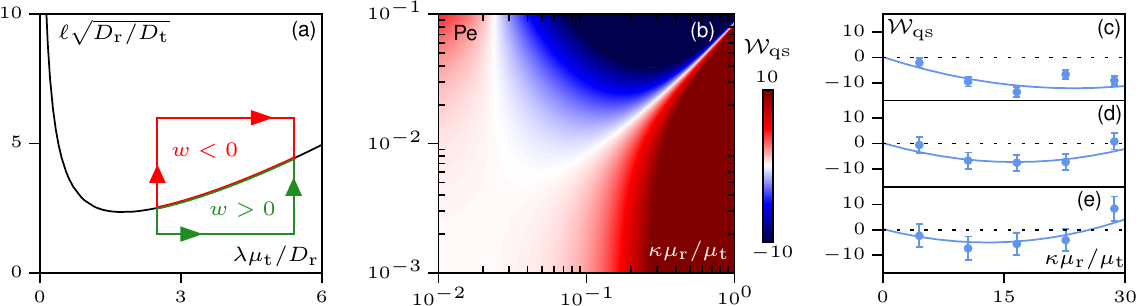}
	\caption{\label{fig2}
		(a)~The square protocol of scaled volume $\ell \sqrt{D_{\rm r}/D_{\rm t}}$ and scaled stiffness $\lambda\mu_{\rm t}/D_{\rm r}$ splits into sub-cycles with opposite directions when it crosses the black solid line $w(\lambda, \ell)=0$, where $w$ obeys ${\cal W}_{\rm qs} = N \iint w(\lambda, \ell) {\rm d}\lambda{\rm d}\ell$.
		(b)~Average quasistatic work ${\cal W}_{\rm qs}$ produced with a clockwise square protocol as a function of the P\'eclet number $\rm Pe$ and of the scaled particle anisotropy $\kappa \mu_{\rm r}/\mu_{\rm t}$. Blue and red regions respectively refer to work extraction for clockwise and counter-clockwise cycles.
		(c-e)~Numerical simulations and corresponding analytical predictions, respectively shown in points and solid lines, illustrate the non-monotonic behavior of ${\cal W}_{\rm qs}$ with $\kappa$. Parameters: ${\rm Pe} =0.2$ (c), $0.067$ (d), and $0.033$ (e). 
		Simulation details in~\cite{Supplemental}.
	}
\end{figure*}

The pressure was already computed in~\cite{Solon2015a} as
\begin{equation}\label{eq:pressure}
	P = \frac{\rho(\ell,\lambda) D_{\rm t}}{\mu_{\rm t}} \bigg[ 1 + {\rm Pe}\, \phi\bigg(\frac{\lambda\kappa\mu_{\rm r}}{D_{\rm r}}\bigg) \bigg] ,
	\quad
	\phi(z) = \frac{1-{\rm e}^{-z}}{z} .
\end{equation}
We evaluate the confining energy from the Fokker-Planck equation associated with the dynamics~\eqref{eq:dyn}, yielding~\cite{Supplemental}
\begin{equation}\label{eq:confining}
	\begin{aligned}
		N\langle u_{\rm t}\rangle_{\rm s} &= \frac{D_{\rm t}\big[N - \ell\,\rho(\ell,\lambda) \big]}{2\mu_{\rm t}} \bigg[ 1 + {\rm Pe}\, \psi\bigg(\frac{\lambda\kappa\mu_{\rm r}}{D_{\rm r}}, \frac{\lambda\mu_{\rm t}}{D_{\rm r}}\bigg) \bigg] ,
		\\
		\psi(z,z') &= \frac{\phi(z) \big[1-\chi(z)\big]}{1+z'\phi(z)} ,
		\quad
		\chi(z) = \frac{I_1(z/2)}{I_0(z/2)} ,
	\end{aligned}
\end{equation}
where $I_n$ is the modified Bessel function of the first kind. The confining energy~\eqref{eq:confining} follows the equipartition theorem at equilibrium (${\rm Pe}=0$), and the nonequilibrium correction for ${\rm Pe}> 0$ yields a dependence on orientation parameters. Since orientations follow an equilibrium dynamics in the wall regions, we get the aligning energy by averaging over the Boltzmann weight ${\rm e}^{-\mu_{\rm r}u_{\rm r}/D_{\rm r}}$, yielding
\begin{equation}\label{eq:potential}
	N\langle u_{\rm r}\rangle_{\rm s} = \frac{\lambda \kappa\big[\ell\,\rho(\ell,\lambda)-N\big]}{2} \,\chi\bigg(\frac{\lambda\kappa\mu_{\rm r}}{D_{\rm r}}\bigg) .
\end{equation}
Combining~(\ref{eq:work_cycle}-\ref{eq:potential}), the work then follows as
\begin{equation}\label{eq:work_fin}
	\begin{aligned}
		{\cal W}_{\rm qs} &= \frac{v^2}{2\mu_{\rm t}D_{\rm r}} \oint_{\partial\Sigma} \bigg\{ - \rho(\ell,\lambda) \,\phi\bigg(\frac{\lambda\kappa\mu_{\rm r}}{D_{\rm r}}\bigg) {\rm d}\ell
		\\
		&\quad + \frac{1}{2}\big[N - \ell\,\rho(\ell,\lambda) \big] \,\psi\bigg(\frac{\lambda\kappa\mu_{\rm r}}{D_{\rm r}}, \frac{\lambda\mu_{\rm t}}{D_{\rm r}}\bigg) \frac{{\rm d}\lambda}{\lambda} \bigg\}
		\\
		&\quad + \frac{\kappa}{2} \oint_{\partial\Sigma} \big[\ell\,\rho(\ell,\lambda)-N \big] \,\chi\bigg(\frac{\lambda\kappa\mu_{\rm r}}{D_{\rm r}}\bigg) {\rm d}\lambda ,
	\end{aligned}
\end{equation}
where we have identified a boundary term of the form $\oint{\rm d}\ln\rho(\ell,\lambda)=0$. The three lines in~\eqref{eq:work_fin} correspond respectively to contributions from the pressure as the volume changes, and from the confining and aligning potentials as the wall stiffness changes.


The work ${\cal W}_{\rm qs}$ can take either signs depending on whether the cycle operates clockwise or counter-clockwise in the space of volume and stiffness. To determine the appropriate direction for extracting work (${\cal W}_{\rm qs}<0$), it is sufficient to know the sign of the surface integrand $w$ defined by
\begin{equation}
		{\cal W}_{\rm qs} = \,N\iint_\Sigma w(\lambda, \ell) \,{\rm d}\lambda{\rm d}\ell ,
		\quad
		w(\lambda, \ell) = \frac{\partial_\lambda P}{N} + \frac{\partial_\ell\langle u_{\rm tot}\rangle_{\rm s}}{\lambda} ,
\end{equation}
where here the cycle is clockwise in the $\{\lambda, \ell\}$ plane, see Fig.~\ref{fig2}(a). For a given range of volume $\ell$ and stiffness $\lambda$, the protocol which realizes maximal work is a square running clockwise (counter-clockwise) for $w<0$ ($w>0$) when the sign of $w$ is fixed within the whole surface $\Sigma$. In contrast, when $\Sigma$ intersects the null line $w=0$, the optimal protocol no longer corresponds to $\ell$ and $\lambda$ varying independently. Instead, one has now to make a choice between the sub-protocols which enclose the parts where $w$ has a constant sign. In particular, when these sub-protocols enclose exactly opposite values of $w$, the work of the associated square cycle vanishes.

Changing internal parameters affects the shape of the null line, whose coordinates follow directly from~(\ref{eq:rho}-\ref{eq:potential}), which can yield a transition between having either clockwise or counter-clockwise cycles to extract work (${\cal W}_{\rm qs}<0$). We recapitulate this transition in the diagram of particle anisotropy $\kappa$ and P\'eclet number $\rm Pe$ shown in Fig.~\ref{fig2}(b). At fixed $\rm Pe$, the work has a non-monotonic dependence on $\kappa$, as confirmed by numerics in Figs.~\ref{fig2}(c-e). When ${\rm Pe}\ll1$ or $\lambda\kappa\mu_{\rm r}\gg D_{\rm r}$, the contribution of $\langle u_{\rm r}\rangle_{\rm s}$ to the work, given by the third term in~\eqref{eq:work_fin}, dominates others. In this regime, the pressure follows an equation of state, which does not preclude extracting work from orientational degree of freedoms. In practice, increasing (decreasing) the stiffness $\lambda$ lowers (elevates) the bottom of the aligning potential $u_{\rm r}$, hence extracting (providing) energy from (to) the particles. Since more particles align with the walls at small volume, the protocol should compress and expand respectively at small and large stiffness in order to extract more energy when increasing $\lambda$ than the one provided when decreasing $\lambda$. This corresponds to counter-clockwise cycle, see red regions in Fig.~\ref{fig2}(b).

\begin{figure}
	\centering
	\includegraphics[width=\linewidth]{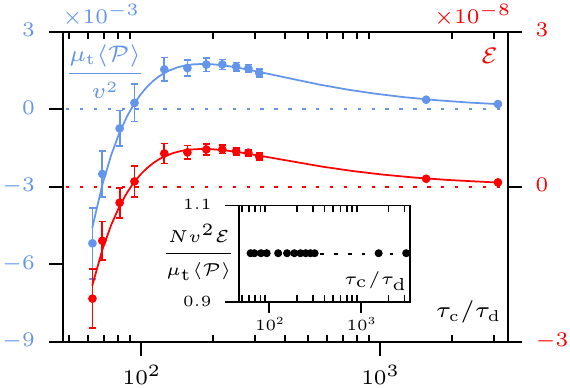}
	\caption{\label{fig3}
		Scaled power $\mu_{\rm t}\langle{\cal P}\rangle/v^2$ and efficiency $\cal E$ as functions of the scaled cycle time $\tau_{\rm c}/ \tau_{\rm d}$, where $\tau_{\rm d} = \ell^2/D_{\rm t}$. They reach a peak value at finite cycle time, and follow the proportionality relation ${\cal E} = \mu_{\rm t}\langle{\cal P}\rangle/(Nv^2)$ shown in the inset. The solid lines refer to the best fits $\langle{\cal P}\rangle = ({\cal W}_{\rm qs}/\tau_{\rm c}) (\tau_{\rm r}/\tau_{\rm c} - 1)$ where $\tau_{\rm r}$ is the only free parameter.
		Simulation details in~\cite{Supplemental}.
	}
\end{figure}

We now turn to discussing finite-time protocols where volume and stiffness no longer vary slowly compared with particle relaxation. Though the quasistatic case is useful to build intuition on how to operate the cycle, it has only a limited application since the power extracted per cycle, $\cal P$, vanishes on average at large cycle time $\tau_{\rm c}$:
\begin{equation}
	{\cal P} = - \frac{\cal W}{\tau_{\rm c}} ,
	\quad
	{\cal W} = N \int_0^{\tau_{\rm c}} \bigg( \frac{\partial u_{\rm tot}}{\partial\ell} \,\dot\ell + \frac{\partial u_{\rm tot}}{\partial\lambda} \,\dot\lambda \bigg) {\rm d} t ,
\end{equation}
where $\cal W$ is the finite-time work. At small cycle time, the cycle does not extract work ($\langle{\cal W}\rangle>0$), and the average power reaches a peak value for intermediate cycle time, as shown in Fig.~\ref{fig3}. In practice, our numerical data are well fitted by $\langle{\cal P}\rangle = ({\cal W}_{\rm qs}/\tau_{\rm c}) (\tau_{\rm r}/\tau_{\rm c} - 1)$ where $\tau_{\rm r}$ is the only free parameter, as expected from linear response~\cite{Seifert2008, Esposito2010}.

Building on thermodynamic uncertainty relations~\cite{Pietzonka2016a, Gingrich2016}, recent works have put forward a generic relation between the power $\cal P$ and the heat $\cal Q$~\cite{Pietzonka2018, Koyuk2019}:
\begin{equation}\label{eq:tur}
	\frac{1}{\big\langle ({\cal P} - \langle{\cal P}\rangle)^2 \rangle} \Bigg[ \big\langle{\cal P}\big\rangle + \tau_{\rm c}\, \frac{{\rm d}\big\langle{\cal P}\big\rangle}{{\rm d}\tau_{\rm c}} \Bigg]^2 \leq \frac{\big\langle{\cal Q}\big\rangle}{2T} .
\end{equation}
It holds for any cyclic protocol independently of the microscopic details, hence being valid for both thermal and active cycles. As a straightforward extension of the thermal case~\cite{Sekimoto1998, Seifert2012}, the heat of active cycles equals the work done by the particles on the thermostat, provided that the forces $\{\dot{\bf r}_i/\mu_{\rm t}, \dot\theta_i/\mu_{\rm r}\}$ and $\{\sqrt{2 D_{\rm t}}{\boldsymbol\xi}_i/\mu_{\rm t}, \sqrt{2D_{\rm r}}\eta_i/\mu_{\rm r}\}$ indeed stem from the surrounding solvent, respectively as damping and thermal fluctuating contributions:
\begin{equation}\label{eq:heat}
	{\cal Q} = \sum_{i=1}^N \int_0^{\tau_{\rm c}} \bigg[ \frac{\dot{\bf r}_i}{\mu_{\rm t}} \cdot \big( \dot{\bf r}_i - \sqrt{2D_{\rm t}}{\boldsymbol\xi}_i \big) +\frac{\dot\theta_i}{\mu_{\rm r}} \big(\dot\theta_i - \sqrt{2D_{\rm r}}\eta_i\big) \bigg] \,{\rm d}t ,
\end{equation}
where the integral is interpreted in Stratonovich sense. The average heat $\langle{\cal Q}\rangle$ is always positive, as a signature of the irreversibility of the dynamics~\cite{Speck2016, Ahmed2016, Nardini2016, Cagnetta2017, Seifert2018, Bo2019}.

Substituting the dynamics~\eqref{eq:dyn} in~\eqref{eq:heat}, we get
\begin{equation}\label{eq:law}
	\big\langle{\cal Q}\big\rangle = \big\langle{\cal W}\big\rangle + \frac{v}{\mu_{\rm t}} \sum_{i=1}^N \int_0^{\tau_{\rm c}} \big\langle \dot{\bf r}_i \cdot {\bf e}_i \big\rangle \,{\rm d}t ,
\end{equation}
where we have used the chain rule $\dot u_{\rm tot} = \big[ \dot\ell \partial_\ell + \dot\lambda \partial_\lambda \big]  u_{\rm tot} + \sum_i \big[ \dot\theta_i \partial_{\theta_i}	 + \dot{\bf r}_i\cdot\nabla_i \big] u_{\rm tot} $ and the stationarity condition $\langle u_{\rm tot}(0)\rangle = \langle u_{\rm tot}(\tau_{\rm c})\rangle$. The expression of average heat in~\eqref{eq:law} clearly differs from the standard first law of thermodynamics: This is at variance with other studies of active cycles which rather define heat by enforcing a relation in terms of work and potential energy as in thermal systems~\cite{Zakine2017, Saha2018, Saha2020, Kroy2020, Lee2020}. Importantly, our definition captures the fact that heat is dissipated even when the potential is static ($\langle{\cal W}\rangle=0$), which stems from the microscopic self-propulsion $v{\bf e}_i$. Provided that most particles evolve in the bulk region without being affected by the confining potential $u_{\rm t}$, the average heat can be simplified using $\sum_i\langle \dot{\bf r}_i\cdot{\bf e}_i \rangle = N v - \mu_{\rm t} \sum_i \langle {\bf e}_i \cdot \nabla_i u_{\rm t} \rangle \approx N v $, yielding
\begin{equation}\label{eq:heat_bis}
	\big\langle{\cal Q}\big\rangle \approx \tau_{\rm c} \,\big[ N v^2/\mu_{\rm t} - \big\langle{\cal P}\big\rangle \big] .
\end{equation}
It follows that~\eqref{eq:tur} reduces to a constraint only between the average and the variance of the power for any cycle time. In particular, at maximum average power (${\rm d}\langle{\cal P}\rangle/{\rm d}\tau_{\rm c}=0$), we get
\begin{equation}\label{eq:tur_bis}
	\frac{\big\langle{\cal P}\big\rangle^2}{\big\langle ({\cal P} - \langle{\cal P}\rangle)^2 \rangle} \leq \frac{Nv^2/\mu_{\rm t} - \big\langle{\cal P}\big\rangle}{2T\tau_{\rm c}} .
\end{equation}
The uncertainty relation~\eqref{eq:tur_bis} remains valid beyond the specific case of varying volume and stiffness as long as (i)~the protocol consists in changing only the potential at boundaries, and (ii)~interactions between particles are neglected.

To characterize further the engine performances, we consider the cycle efficiency $\cal E$. Following standard definitions for monothermal protocols~\cite{Prost1999, Pietzonka2016, Pietzonka2019}, it reads
\begin{equation}
	{\cal E} = \frac{\big\langle{\cal W}\big\rangle}{\big\langle{\cal W}\big\rangle-\big\langle{\cal Q}\big\rangle} \leq 1 ,
\end{equation}
from which, by using~\eqref{eq:heat_bis}, we deduce
\begin{equation}\label{eq:efficiency}
	{\cal E} \approx \frac{\mu_{\rm t}\big\langle{\cal P}\big\rangle}{Nv^2} .
\end{equation}
Considering a square protocol where $\ell$ and $\lambda$ vary linearly in time, the efficiency and power measured numerically indeed confirm~\eqref{eq:efficiency}, as shown in Fig.~\ref{fig3}. The proportionality relation~\eqref{eq:efficiency} assumes that the bulk region is large compared with the wall penetration length, which typically leads to a modest efficiency: Most particles dissipate energy in the bulk without contributing to the work produced at boundaries. Conversely, reducing the relative bulk size compared with the typical penetration length within the walls should increase the efficiency, though the assumption of flat bulk profile, used when deriving quasistatic work, can break down in this regime. Note that increasing the system size along the direction parallel to walls also leads to higher efficiency.

Importantly, the efficiency is maximum at finite cycle time, in contrast with thermal engines where quasistatic protocols always realize maximal efficiency~\cite{Carnot}. This is because active particles dissipate energy even when the potential is static, so that the energy cost increases with cycle time and thus one cannot afford to operate the cycle infinitely slowly. In practice, the efficiency~\eqref{eq:efficiency} only accounts for the transfer of energy from particle motion to work extraction: It deliberatly discards energy exchanges at the basis of the microscopic self-propulsion consuming fuel supply. Provided that fuel consumption operates faster than the typical relaxation of positions and orientations, it should not be affected by the cycle time of external protocols. Then, the cycle still achieves maximum efficiency at finite time even when accounting for such a consumption.


In this paper, we have provided a consistent thermodynamic framework for cycles operating with active matter. The approach for identifying the appropriate cycle direction, which relies on evaluating the deviation from Boltzmann statistics $w(\alpha_1, \alpha_2)$ in~\eqref{eq:work_green}, carries over beyond our case study: It gives a recipe for evaluating and comparing the properties of various cycles~\cite{Zakine2017, Martin2018, Saha2018, Saha2020, Kroy2020, Lee2020}. Importantly, we demonstrate that one can extract work without changing any property of active particles, since it is sufficient to control only the potential at boundaries. Thus, our work offers guidelines for future experiments of active engines, based on manipulating either colloidal~\cite{Sood2016} or macroscopic~\cite{Dauchot2017} active particles, where the properties of confining walls can be varied externally. A potential realization of soft walls consists in adding polymer brushes on surfaces, whose extension is controlled for instance by ionic concentration~\cite{Milner1991, Azzaroni2012}. Based on our framework, it would be interesting to propose ideal protocols which bound the cycle performances, analogously to the Carnot cycle for thermal engines~\cite{Carnot}. In stark contrast with thermal cycles, which entail a trade-off between power and efficiency~\cite{Seifert2008, Esposito2010, Shiraishi2016, Pietzonka2018}, our cycles reach simultaneously maximum power and efficiency. The challenge is then to find optimal protocols, where the control parameters have potentially a complex time dependence beyond linear behavior, to increase efficiency and power at finite cycle time.

\acknowledgements{The authors acknowledge insightful discussions with Robert L. Jack, Timur Koyuk and Patrick Pietzonka. Work funded in part by the European Research Council under the EU’s Horizon 2020 Programme, grant number 740269. \'EF benefits from an Oppenheimer Research Fellowship from the University of Cambridge, and a Junior Research Fellowship from St Catharine's College. MEC is funded by the Royal Society.}


\bibliography{References.bib}

\end{document}